\documentclass[sn-mathphys]{sn-jnl}

\usepackage{array}
\usepackage{colortbl}
\usepackage{lineno}
\usepackage{caption}
\usepackage{subcaption}
\usepackage{graphicx}
\newcommand{\upcite}[1]{\textsuperscript{\citenum{#1}}}

\makeatletter
\newenvironment{figurehere}
{
    \def\@captype{figure}
}

{}
\makeatletter

\usepackage{makecell}
\usepackage{pifont}
\usepackage[version=4]{mhchem}
\usepackage{float}
\usepackage{multicol}

\usepackage{geometry}

\jyear{2023}%

\theoremstyle{thmstyleone}%
\theoremstyle{thmstyletwo}%

\theoremstyle{thmstylethree}%

\begin{document}

\title[HyperspecI Sensor]{A broadband hyperspectral image sensor with high spatio-temporal resolution}

\author*[]{\fnm{Liheng} \sur{Bian*}}\email{bian@bit.edu.cn}
\equalcont{These authors contributed equally to this work.}

\author[]{\fnm{Zhen} \sur{Wang}}
\equalcont{These authors contributed equally to this work.}

\author[]{\fnm{Yuzhe} \sur{Zhang}}
\equalcont{These authors contributed equally to this work.}
\author[]{\fnm{Lianjie} \sur{Li}}
\author[]{\fnm{Yinuo} \sur{Zhang}}
\author[]{\fnm{Chen} \sur{Yang}}
\author[]{\fnm{Wen} \sur{Fang}}
\author[]{\fnm{Jiajun} \sur{Zhao}}
\author[]{\fnm{Chunli} \sur{Zhu}}
\author[]{\fnm{Qinghao} \sur{Meng}}
\author[]{\fnm{Xuan} \sur{Peng}}
\author*[]{\fnm{Jun} \sur{Zhang*}}\email{zhjun@bit.edu.cn}

\affil[]{State Key Laboratory of CNS/ATM \& MIIT Key Laboratory of Complex-field Intelligent Sensing, Beijing Institute of Technology, Beijing 100081, China}

\abstract
Hyperspectral imaging provides high-dimensional spatial-temporal-spectral information revealing intrinsic matter characteristics\upcite{landgrebe2002hyperspectral, li2019deep, backman2000detection, hadoux2019non, mehl2004development}. Here we report an on-chip computational hyperspectral imaging framework with high spatial and temporal resolution. By integrating different broadband modulation materials on the image sensor chip, the target spectral information is non-uniformly and intrinsically coupled on each pixel with high light throughput. Using intelligent reconstruction algorithms, multi-channel images can be recovered from each frame, realizing real-time hyperspectral imaging. Following such a framework, we for the first time fabricated a broadband VIS-NIR (400-1700 nm) hyperspectral imaging sensor using photolithography, with an average light throughput of 74.8$\%$ and 96 wavelength channels. The demonstrated resolution is 1,024$\times$1,024 pixels at 124 fps. 
We demonstrated its wide applications including chlorophyll and sugar quantification for intelligent agriculture, blood oxygen and water quality monitoring for human health, textile classification and apple bruise detection for industrial automation, and remote lunar detection for astronomy. The integrated hyperspectral image sensor weighs only tens of grams, and can be assembled on various resource-limited platforms or equipped with off-the-shelf optical systems. The technique transforms the challenge of high-dimensional imaging from a high-cost manufacturing and cumbersome system to one that is solvable through on-chip compression and agile computation.

\keywords{hyperspectral imaging, integrated sensor, VIS-NIR, on-chip modulation, high resolution, low cost}

\maketitle

\section{Introduction}\label{sec1}

Hyperspectral imaging captures spatial, temporal, and spectral information of the physical world, characterizing the intrinsic optical properties of each location\upcite{landgrebe2002hyperspectral}. Compared to multispectral imaging, hyperspectral imaging acquires a significantly large number of wavelength channels ranging from tens to hundreds, and maintains a superior spatial mapping capability compared to spectrometry\upcite{yang2019single}. Such great high-dimensional information enables precise distinguishment of different materials with similar colors, empowering more intelligent inspection than human vision with higher spectral resolution and wider spectral range. With such advantages, hyperspectral imaging has been widely applied in various fields such as remote sensing, machine vision, agricultural analysis, medical diagnostics, and scientific monitoring\upcite{li2019deep,backman2000detection,hadoux2019non,mehl2004development}. 

The most severe challenge to realizing hyperspectral imaging is how to efficiently acquire the dense spatial-spectral data cubes. 
Most of the existing hyperspectral imaging systems employ individual optical elements (like prism, grating, or spectral filters) and mechanical components to scan hyperspectral cubes in the spatial or spectral dimension\upcite{green1998imaging}. However, these systems typically suffer from drawbacks such as large size, heavy weight, high cost, and time-consuming operation, which limit their widespread application. 
As advanced by the compressive sensing theory and computational photography technique\upcite{pian2017compressive}, various computational snapshot hyperspectral imaging techniques have been developed such as computed-tomography imaging system (CTIS)\upcite{descour1995computed} and coded aperture snapshot spectral imaging (CASSI)\upcite{wagadarikar2008single}, which encode multi-dimensional hyperspectral information into single-shot measurements, and then decode the data cube via compressive sensing or deep learning algorithms. 
Although these systems effectively improve temporal resolution, they still require individual optical elements for explicit light modulation that takes a heavy load for lightweight integration\upcite{arguello2014colored}.

Aiming for integrated hyperspectral imaging, there have been several trials for on-chip acquisition. The most intuitive routine is to extend the classic Bayer pattern of RGB color cameras by introducing more narrow-band filters that has led commercial multispectral imaging sensors\upcite{geelen2014compact}. However, this technique suffers from a significant tradeoff between spatial and spectral resolution, and it also wastes the most light throughput due to narrow-band filtering. Benefiting from finely tunable spectrum filtering ability, nano-fabricated metasurface\upcite{yesilkoy2019ultrasensitive,faraji2019hyperspectral,xiong2022dynamic, he2024meta}, photonic crystal slabs arrays\upcite{wang2019single}, and Fabry-Pérot filters\upcite{yako2023video} have also been employed for spectral modulation in a certain spectral range. Experimentally, most of these existing prototypes cover $\sim$200 nm in the visible range\upcite{yako2023video}, with only $\sim$20 channels.
Recently, scattering media have been utilized for compact lensless hyperspectral imaging systems, building on its spatial multiplexing and point spread function properties\upcite{kim2023aperture,redding2013compact,monakhova2020spectral,jeon2019compact}.
Despite the above various on-chip techniques, most of them suffer from narrow spectral range, low light throughput, and the intrinsic tradeoff between spatial and spectral resolution. 
The comprehensive performance comparison of different techniques is referred to Extended Data Tab. \ref{Extended_Table_Comparison}.

In this work, we report an on-chip computational hyperspectral imaging sensor, termed the HyperspecI sensor, with its comprehensive framework of hardware fabrication, optical calibration, and computational reconstruction. First, to acquire both spatial and spectral information effectively, we developed a broadband multispectral filter array (BMSFA) fabrication technique using photolithography. The BMSFA is composed of different broadband spectral modulation materials at different spatial locations. 
Different from the common narrow-band filters, the BMSFA can modulate incident light at the entire wide spectral range, allowing much higher light throughput that benefits low-light and long-distance imaging applications. 
The modulated information is then intrinsically compressed and acquired by the underlying broadband monochrome sensor chip, allowing for spatial-spectral compression with full temporal resolution.
Second, to efficiently restore hyperspectral data cubes from the BMSFA compressed measurements, we derived a light-weight and high-performance neural network (SRNet), which has stronger feature extraction and prior modeling ability.
Consequently, we can reconstruct hyperspectral images (HSIs) with high spatial and spectral resolution from each frame, realizing high-throughput real-time hyperspectral imaging.

Following the above framework, we for the first time fabricated two VIS-NIR hyperspectral image sensors (HyperspecI-V1 and HyperspecI-V2). The spectral response range covers 400-1000 nm and 400-1700 nm respectively, with an average light throughput of 71.8$\%$ and 74.8$\%$. In low-light conditions, the HyperspecI sensors outperform mosaic multispectral cameras and scanning hyperspectral systems a lot, as validated in Fig. \ref{fig:3}. 
The average spectral resolution achieves 2.65 nm for HyperspecI-V1 and 8.53 nm for HyperspecI-V2. 
As for hyperspectral imaging, the HyperspecI-V1 sensor produces 61 channels within the 400-1000 nm range, each with 2,048 $\times$ 2,048 pixels at 47 fps. The HyperspecI-V2 sensor demonstrates 96 wavelength channels with a 10 nm interval within the range of 400-1000 nm and a 20 nm interval within the range of 1000-1700 nm. Each channel consists of 1,024 $\times$ 1,024 pixels at 124 fps.
A more detailed performance illustration is referred to Extended Data Tab. \ref{Extended_Table_Comparison}.

To demonstrate the practical capabilities and wide application potentials of the HyperspecI sensors, we conducted soil plant analysis development (SPAD) and soluble solid content (SSC) evaluation for intelligent agriculture, blood oxygen and water quality monitoring for human health, textile classification and apple bruise detection for industrial automation, and remote lunar detection for astronomy. These applications demonstrate the irreplaceable advantage of high signal-to-noise ratio (SNR), high resolution, ultra-broadband, and dynamic hyperspectral imaging capabilities of our HyperspecI technique, providing unique benefits in such conditions under low light, targeting dynamic scenes, and detecting small or remote targets that are unattainable using other techniques. Furthermore, the compact size, lightweight, and high integration level of HyperspecI make it suitable for deployment on platforms with limited payload capacity. We anticipate that this scheme may open a new venue for the next-generation image sensor of higher information dimension, higher imaging resolution, and higher degree of intelligence.

\section{Principle of HyperspecI}\label{sec2}

The HyperspecI sensor consists of two main components: a broadband multispectral filter array (BMSFA) and a broadband monochrome image sensor chip (as shown in Fig. \ref{fig:1}a). The BMSFA encodes the high-dimensional hyperspectral information of the target scene in the spectral domain, and the underlying image sensor chip acquires the coupled two-dimensional measurements (as shown in Fig. \ref{fig:1}d). Through a hybrid neural network SRNet, multi-channel hyperspectral images (HSIs) can be reconstructed from each frame with high fidelity and efficiency (Extended Data Fig. \ref{ED_Fig6_SRNet}a-b, Supplementary Section 5).

We developed a photolithography technique to fabricate BMSFA.
First, we prepared broadband materials based on organic materials with different spectral responses. Then, by coupling the broadband materials with the negative photoresist, we fabricated lithography-capable broadband spectral modulation materials (as shown in Fig. \ref{fig:1}c). The materials were selected for optimal broadband spectral modulation characteristics (Extended Data Fig. \ref{ED_Fig9_ModulationMaterials}, Supplementary Section 2). Then, through an improved photolithography process, we solidified the spectral modulation materials on a high transmission quartz substrate following the pre-designed photomask, forming the BMSFA (Fig. \ref{fig:1}b, Extended Data Fig. \ref{ED_Fig7_BMSFAFabrication}). The photolithography process includes a series of steps, including photomask design, substrate preparation, photoresist coating, soft bake, UV exposure, post-exposure bake, development, and hard bake (Supplementary Section 3). 
To meet the demands for different spectral imaging ranges, we designed and prepared BMSFAs with different material systems and spatial arrangements. We integrated the fabricated BMSFAs with CMOS and InGaAs image sensor chips respectively, as shown in Fig. \ref{fig:1}a and Extended Data Fig. \ref{ED_Fig7_BMSFAFabrication}b-c.
 
Figure \ref{fig:1}e presents exemplar hyperspectral imaging results, demonstrating that the HyperspecI sensors can acquire rich spatial details and maintain high spectral accuracy across a wide spectral range. The comprehensive imaging results of more channels and scenes are listed in Extended Data Fig. \ref{ED_Fig3_SpectralImage}. Besides, to demonstrate its high accuracy and efficiency for hyperspectral image reconstruction, we compared the reported SRNet with the existing state-of-the-art model-based and deep learning-based algorithms, which indicates that our SRNet model outperforms others in terms of both accuracy and efficiency metrics (Supplementary Section 5.3). 
Using the HyperspecI sensors, Figure \ref{fig:1}f shows the structure of the collected image and video dataset, which may provide essential help for further hyperspectral imaging and sensing studies.

\section{Performance of HyperspecI}\label{sec3}

We conducted a series of experiments to validate both the quantitative and qualitative performance of the HyperspecI sensors. First, we examined the spectral and spatial resolution of the HyperspecI sensors. 
Figure \ref{fig:2}a and Extended Data Fig. \ref{ED_Fig3_SpectralImage} visualize the reconstructed HSIs in the synthesized RGB format. 
We also compared the reconstructed spectra with corresponding ground truth collected by commercial spectrometers (Ocean Optics USB 2000+ and NIR-Quest 512) at the locations indicated by yellow markers in the synthesized RGB images. 
We presented the reconstruction results of monochromatic light with an interval of 0.2 nm, and compared the reconstruction results of our HyperspecI sensors with commercial spectrometer under single-peak monochromatic light (Fig. \ref{fig:2}b).
The full width at half maximum (FWHM) of the monochromatic light is 2 nm. The average spectral resolution of HyperspecI-V1 and HyperspecI-V2 sensors are 2.65 nm and 8.53 nm, respectively (Fig. \ref{fig:2}b (iii-iv)). Beside, we also employed double-peak monochromatic light (FWHM 2 nm) to calibrate the spectral resolving ability of our sensors based on the Rayleigh criterion.
The results demonstrate the average resolvable double-peak distance of HyperspecI-V1 and HyperspecI-V2 reaching 3.23 nm and 9.76 nm, respectively.
Second, to evaluate the spatial resolution, we acquired images of the USAF 1951 spatial resolution test chart using our HyperspecI sensors and corresponding monochrome cameras (with the same sensor chips and lens configuration), respectively. We presented the HyperspecI-V1 results in Fig. \ref{fig:2}c as a demonstration. The results show that the HyperspecI sensor can distinguish the fourth element of the third group, where the width of the three lines is about 0.26 mm on the chart and occupies 9 pixels of the image, resulting in a spatial resolution of 11.31 lines per millimeter, which is comparable to the commercial monochrome camera.
Furthermore, light throughput comparison among several representative hyperspectral imaging techniques is depicted in Fig. \ref{fig:2}d. The comparison illustrates that the average light throughput is 71.8$\%$ of HyperspecI-V1 and 74.8$\%$ of HyperspecI-V2, which are much higher than that of common RGB color cameras ($\textless 30\%$), mosaic multispectral cameras ($\textless 10\%$), and CASSI systems ($\textless 50\%$) (Supplementary Section 6.5). 

We conducted an imaging experiment on small point targets with varying sizes (Fig. \ref{fig:3}a). The results indicate that the HyperspecI sensor can achieve stable and accurate spectral reconstruction even when the radius of the targets is smaller than a superpixel.
In Fig. \ref{fig:3}b, we also compared the hyperspectral imaging performance among our HyperspecI sensor, a commercial mosaic multispectral camera (Silios, CMS-C), and a scanning hyperspectral camera (FigSpec, FS-23) in low-light conditions. The light source is Thorlabs SLS302 with an illuminance level of 290 lux. 
These experiments demonstrate that our sensor exhibits superior hyperspectral imaging quality in low-light environments, attributed to its higher light throughput and SNR. The superiority is further illustrated by the remote lunar detection experiment presented in Extended Data Fig. \ref{ED_Fig1_Moon}. 
We further demonstrated the real-time imaging performance of our HyperspecI-V1 sensor at a frame rate of 47 fps (Fig. \ref{fig:3}c). 
As a comparison, we presented the imaging results using the scanning hyperspectral imaging camera. The result comparison validates that our HyperspecI sensor achieves a full temporal resolution of the underlying image sensor chip for dynamic imaging at a high frame rate, while the traditional scanning hyperspectral cameras are unable to capture dynamic scenes (Supplementary Section 6.2). 

The above experiments have demonstrated a broad spectral range, high spatial resolution, high spectral accuracy, high light throughput, and real-time frame rate of our HyperspecI sensors. Besides, we have further studied the HyperspecI sensors regarding SNR (Supplementary Section 6.1), noise resistance (Supplementary Section 6.3), dynamic range (Supplementary Section 6.4), and thermal stability (Extended Data Fig. \ref{ED_Fig4_ThermalStability}, Supplementary Section 6.7).

\section{Application for intelligent agriculture}\label{sec4}

Effective detection of target components is imperative to improve crop management strategies\upcite{cortes2019monitoring}. The soil plant analysis development (SPAD) index, highly correlated with the chlorophyll content\upcite{limantara2015analysis}, is pivotal in assessing plant physiology. Similarly, the soluble solid content (SSC) is an important indicator for fruit quality assessment and determination of harvest time\upcite{li2022calibration}. However, conventional SPAD and SSC measurements involve destructive sampling, which is complicated and time-consuming. Advancements in molecular spectroscopy, coupled with chemometric techniques, have popularized VIS-NIR spectroscopy as a non-destructive alternative for internal quality assessment\upcite{ma2021rapid}. To demonstrate the applicability of the HyperspecI sensor in intelligent agriculture, we developed a prototype for non-destructive SPAD and SSC measurements (Fig. \ref{fig:4}a). 

Figure \ref{fig:4}b illustrates SPAD detection principles based on the Lambert-Beer law, utilizing the HyperspecI sensor to acquire transmission spectra of 200 leaves. The values at the characteristic peaks (660 nm and 720 nm) were utilized to establish the regression model. Validation with the additional 20 leaves resulted in high precision with the root mean square error (RMSE) of 1.0532 and a relative error of 3.73\%. Figure \ref{fig:4}c outlines the non-destructive SSC detection procedure in apples. Spectral curves reveal peaks and troughs indicative of various apple characteristics. Our partial least squares regression model accurately predicts SSC, with a correction coefficient of 0.8264 and RMSE of 0.6132\% for the training set, and 0.6162 and 0.7877\% for the test set, respectively. The relative error of the prediction set amounts to 5.30\%. 
Figure \ref{fig:4}d presents RGB and reconstructed hyperspectral images of leaves and apples, highlighting the sensor's potential for agricultural applications. These results emphasize the significant promise of the HyperspecI sensor for non-destructive analysis in intelligent agriculture. For detailed insights, refer to the Supplementary Section 7.

\section{Application for human health}\label{sec5}

The rising attention to health concerns has spurred a proliferation of health monitoring equipment, yet its progress is hampered by limitations in resolution, real-time capabilities, and portability.
To demonstrate the advantages of our HyperspecI in dynamic, high-resolution ability, we conducted experiments on blood oxygen detection and water quality assessment, illustrating its potential for real-time health monitoring as an alternative to traditional bulky and complex equipment.
For blood oxygen saturation monitoring, we developed a prototype device to detect changes in arterial blood absorption at specific wavelengths due to pulsation (Fig. \ref{fig:5}a). 
When the finger under measurement is placed into the device, the transmission spectra are acquired using a broad-spectrum light source and the HyperspecI sensor. By reducing the effective number of pixels in the HyperspecI sensor, we can achieve a collection frame rate of up to 100 Hz.
Subsequently, the acquired data are processed to obtain a series of spectral profiles at a certain area on the finger.
Finally, the pulsatile component (AC) is extracted from the photoplethysmography signal at two characteristic bands (780 nm and 830 nm), which produces blood oxygen saturation (Supplementary Section 8.1). A comparison of measurements between the HyperspecI sensor and a commercial oximeter is presented in Fig. \ref{fig:5}b. 

We further conducted an effluent diffusion monitoring experiment to explore its capability for water quality detection.
During the experiment, two solutions with similar colors but different compositions were rapidly injected into distilled water, and the diffusion process was simultaneously recorded using the HyperspecI sensor and an RGB camera (Fig. \ref{fig:5}c). 
Distinguishing between these two solutions is challenging using RGB images. However, their differentiation becomes straightforward through the disparities in their spectral curves and spectral images at the NIR range (780 nm).
Furthermore, the segmentation results of RGB images and reconstructed hyperspectral images further illustrate the superiority and potential of our HyperspecI in real-time high-resolution spectral imaging and water quality assessment (Fig. \ref{fig:5}d, Supplementary Section 8.2).

\section{Application for industrial automation}\label{sec6}

To demonstrate the near-infrared hyperspectral imaging ability and accuracy of our sensors, we applied them in textile classification and apple bruise detection.
For textile classification, reflectance spectra of textiles were acquired using the HyperspecI sensor (Fig. \ref{fig:6}a).
Previous research\upcite{liu2020qualitative} has shown that characteristic spectral bands of cotton fabrics (at 1220, 1320, and 1480 nm) and polyester fabrics (at 1320, 1420, and 1600 nm) are distinct, facilitating their classification (Fig. \ref{fig:6}b-d).
In our experiment, we prepared 204 samples, including various cotton and polyester fabrics, divided into training (75 cotton and 75 polyester) and testing datasets (27 cotton and 27 polyester).
Given the diverse appearance of these samples, their classification by visual inspection is challenging (Fig. \ref{fig:6}b). 
Subsequently, we employed the Support Vector Machine (SVM) algorithm for automatic fabric categories classification, as shown in Fig.~\ref{fig:6}c. 
For the testing phase, the overall classification accuracy reached 98.15\% (Supplementary Section 9.1).

Apple bruises, often located beneath the skin, are challenging to detect visually, leading to low identification accuracy and efficiency.
Benefiting from the wide spectral range of our HyperspecI, bruised areas exhibit spectral characteristics near wavelengths of 1060 nm, 1260 nm, and 1440 nm due to water absorption of NIR light, which is crucial for invisible bruise detection.
In our experiment, we prepared 224 samples of `Qixia' Fuji apples and used a 30 cm steel pipe to systematically create bruises on random locations of each apple (Fig.~\ref{fig:6}d). We employed our HypersepcI sensor and an RGB camera to acquire hyperspectral and color images of these apples, constructing two separate image datasets (Fig.~\ref{fig:6}e). Each dataset, comprising 224 images, was applied to train a YOLOv5-based detection network, and the rest 40 samples were utilized for testing (Supplementary Section 9.2).
Spectral images were processed to create synthesized color representations, distinctly marking bruised regions for enhanced visualization (Fig.~\ref{fig:6}f).
The detection precision and recall scores on the near-infrared spectral images are significantly higher than those on the RGB images (Fig.~\ref{fig:6}g).
The higher mAP50 and mAP50-95 scores also indicate the effectiveness of utilizing infrared spectral information for apple bruise detection, and further demonstrate that our HyperspecI sensor can capture crucial spectral features of subtle changes in the near-infrared range.

\section{Conclusion and Discussion}\label{sec7}

This work introduces an on-chip hyperspectral image sensor technique, termed HyperspecI, which follows the computational imaging principle to realize integrated and high-throughput hyperspectral imaging. The HyperspecI sensor first acquires encoded hyperspectral information by integrating a broadband multispectral filter array (BMSFA) and a broadband monochrome sensor chip, and then reconstructs hyperspectral images using deep learning. Compared with the classic scanning scheme, the HyperspecI sensor maintains the full temporal resolution of the underlying sensor chip. Compared to the existing snapshot systems, the reported technique demonstrates enhanced integration with lightweight and compact size. Extensive experiments demonstrate the superiority of the HyperspecI sensor on high spatial-spectral-temporal resolution, wide spectral response range, and high light throughput. Such advantages provide irreplaceable benefits in hyperspectral imaging applications such as under low light, targeting dynamic scenes, and detecting unattainable small or remote targets using existing methods. We demonstrated the wide application potentials of the HyperspecI sensor such as intelligent agricultural monitoring and real-time human health monitoring. The different applications validated the versatility, flexibility, and robustness of the HyperspecI technique.

The HyperspecI technique can be further extended.
First, by employing advanced fabrication techniques such as electron beam lithography\upcite{kim2014all}, nanoimprinting, and two-photon polymerization, it is conducive to achieving higher degrees of freedom and precision for BMSFA design and HyperspecI integration.
In addition, considering the excellent compatibility with other materials, the derived BMSFA strategy can be paired with high-performance 2D materials\upcite{yu20172d}, allowing for more precise optical control and enhancing optical performance.
Second, the generalization ability of hyperspectral reconstruction can be further enhanced by training data augmentation, transfer learning, and illumination decomposition, thus helping tackle common challenges such outlier input, metamerism, varying illumination, and so on\upcite{zheng2015illumination}.
Third, the HyperspecI's real-time hyperspectral imaging ability can be combined with heterogeneous detection devices, such as LIDAR and SAR, to achieve multi-source fusion detection\upcite{abdar2021review}. This is critical to realize high-precision sensing and make high-reliability decisions in complex environments.
Fourth, HyperspecI's highly compatible architecture provides off-the-shelf solutions for easy integration with various imaging platforms, thus directly upgrading their sensing dimension and enabling multifunctional applications. For instance, the integration of vibration-coded microlens arrays into the BMSFA may enable high-resolution hyperspectral 3D photography\upcite{wu2022integrated}. The combination with ultrafast imaging systems can realize hyperspectral transient observation\upcite{gao2014single}. 
By further designing BMSFA with multidimensional multiplexing capabilities (such as polarization and phase encoding), large-scale multidimensional imaging can be achieved\upcite{altaqui2021mantis}. When incorporated with fluorescence imaging systems, the fluorescence signals of different dyes can be effectively separated based on spectral characteristics in a snapshot manner, thus improving detection sensitivity and efficiency in biomedicine science\upcite{shi2020pre, wu2021iterative}.
Overall, we believe this work may open a new venue for the next-generation image sensor of higher information dimension, higher imaging resolution, and higher degree of intelligence.

\newpage

\newpage

\begin{figurehere}
	\centering
	\caption{\label{fig:1} {\textbf{Working principle of the HyperspecI technique.}}
		{\textbf{a.}} The HyperspecI sensor consists of a BMSFA mask and a broadband monochrome image sensor chip. The BMSFA consists of a cyclic arrangement of $4 \times 4$ broadband materials for broadband spectral modulation, with each modulation unit 10 \textmu m in size. The BMSFA is cured onto the bare photodiode array surface using SU-8 photoresist.
        {\textbf{b.}} The manufacturing process of BMSFA. We developed a low-cost fabrication strategy to produce BMSFA using photolithography.
        {\textbf{c.}} The transmission spectra of the 16 modulation materials and the coefficient correlation matrix.
        {\textbf{d.}} The imaging principle of the HyperspecI sensor. The light emitted from the target scene is modulated after passing through BMSFA, and then captured by the underlying broadband image sensor chip. The collected compressed data is then input into a reconstruction algorithm to decouple and output HSIs. {\textbf{e.}} The exemplar hyperspectral imaging results of the HyperspecI sensor. {\textbf{f.}} Illustration of the collected large-scale HSI image and video dataset using the HyperspecI sensor.\\}		
\end{figurehere}

\begin{figurehere}
	\centering
	\caption{\label{fig:2} {\textbf{Hyperspectral imaging performance of the HyperspecI sensors.}} 
        {\textbf{a.}} Exemplar hyperspectral imaging results. The reconstructed hyperspectral images are visualized in the synthesized RGB format on the left. The spectral comparison between reconstructed spectra (RS) and ground truth (GT), acquired by the commercial spectrometers, are represented on the right (denoted by solid and dashed lines respectively). 
        {\textbf{b.}} Spectral resolution calibration. (i-ii) Spectral comparison between the HyperspecI sensors (green solid lines) and commercial spectrometer (black dashed lines). The monochromatic light (FWHM 2 nm) was produced by the commercial Omno151 monochromator. (iii-iv) Single-peak and double-peak monochromatic light were used to analyze the spectral resolving ability of our sensors. The average FWHM of reconstructed spectra under single-peak monochromatic light for HyperspecI-V1 and V2 are 2.65 nm and 8.53 nm, respectively. 
        The average resolvable peak distance of reconstructed spectra based on the Rayleigh criterion for HyperspecI-V1 and V2 under double-peak monochromatic light are 3.23 nm and 9.76 nm, respectively.
        {\textbf{c.}} Spatial resolution calibration using the USAF 1951 resolution test chart. The curves of a monochrome camera (red line) and our HyperspecI-V1 sensor (blue line) for elements 1-6 of group 3 are presented. 
        {\textbf{d.}} Light throughput calibration. 
        }
\end{figurehere}

\begin{figurehere}
	\centering
        \caption{\label{fig:3}{\textbf{Hyperspectral imaging performance demonstration on high resolution, high light throughput, and real-time capability.}} 
        {\textbf{a.}} Hyperspectral imaging results of small targets. On the left, we displayed the raw measurements, synthesized RGB images, and hyperspectral images of several exemplar bands. On the right, the comparison is shown among the background spectrum, ground truth, and reconstructed spectra of the small targets, which are marked in the synthesized RGB image with a blue rectangle. 
        {\textbf{b.}} Hyperspectral imaging comparison in low-light conditions. The imaging results of our HyperspecI sensor, a commercial mosaic multispectral camera, and a commercial scanning hyperspectral imaging camera were compared at a fixed exposure time of 1 ms. We presented the synthesized RGB images, exemplar spectral images at 580 nm and 690 nm, and the corresponding normalized data. 
        {\textbf{c.}} Hyperspectral imaging results at video frame rate. On the left, we presented the results at three different time points (0 s, 1 s, and 2 s) while the object was undergoing translational motion at a speed of $\sim$0.5 m/s. 
        On the right, we presented the results at three different time points (0 s, 0.02 s, and 0.04 s) while the object was undergoing rotational motion at a speed of $\sim$6 rad/s. The result comparison was demonstrated using synthesized RGB images and spectral images at 550 nm, 700 nm, and 850 nm.  
        \\}
\end{figurehere}

\begin{figure}[!htbp]
	\centering
	\caption{\label{fig:4}
            {\textbf{Application of the HyperspecI sensor for intelligent agriculture.}} {\textbf{a.}} The prototype using the HyperspecI sensor for agriculture spectra acquisition. It includes two distinct modes: the mode of leaf transmission spectra acquisition and the mode of apple reflectance spectra acquisition.
            {\textbf{b.}} The working principle of measuring soil plant analysis development (SPAD) index, which is used to evaluate the chlorophyll content of leaves, is shown on the left. SPAD evaluation results using the HyperspecI sensor are shown on the right.
            {\textbf{c.}} The working principle of measuring soluble solid content (SSC), used to evaluate apple quality, is shown on the left. The comparison between the measured SSC using a commercial product and the predicted SSC using our HyperspecI sensor and partial least squares (PLS) regression model is shown on the right.        
            {\textbf{d.}} The comparison between RGB images and synthesized RGB images using the reconstructed hyperspectral images. The middle subfigure shows a comparison between the spectra acquired by a commercial spectrometer and the reconstructed spectra at exemplar randomly selected locations.
            }   
\end{figure}

\begin{figure}[!htbp]
	\centering
	\caption{\label{fig:5}{\textbf{Application of the HyperspecI sensor for blood oxygen and water quality monitoring.}}
		{\textbf{a.}} The prototype we built using the HyperspecI sensor for blood oxygen saturation (\ce{SpO_{2}}) monitoring. 
		{\textbf{b.}} (i) The transmission spectra through the finger were obtained at a collection rate of 100 Hz.  (ii) Two photoplethysmography (PPG) signals at 780 nm and 830 nm, corresponding to two bands with different intensities of \ce{HbO_{2}} and Hb absorption. 
        The blood oxygen saturation can be accurately determined by analyzing and calibrating the PPG signals at these two characteristic bands. 
        (iii) Comparative analysis with a commercial oximeter product demonstrates a high level of consistency in the obtained results.
		{\textbf{c.}} Three exemplar frames of HyperspecI measurements demonstrating the solution diffusion process, accompanied by corresponding images captured using an RGB camera. In the petri dish, solution \#1 was positioned in the upper left corner, and solution \#2 was placed in the lower left corner. These two solutions were added to distilled water in the upper right corner. Hyperspectral images acquired by the HyperspecI sensor are presented in synthesized RGB format.
		{\textbf{d.}} Comparison of segmentation maps between an RGB camera (left) and the HyperspecI sensor (right).}
\end{figure}

\begin{figurehere}
	\centering
	\caption{\label{fig:6}
            \textbf{Application of the HyperspecI sensor for textile classification and apple bruise detection. } 
            \textbf{a.} The experiment configuration for the acquisition of fabric spectra. 
            \textbf{b.} Measurements and reconstructed hyperspectral images of textile samples, together with synthesized RGB (sRGB) representations and exemplar hyperspectral images (1220 nm, 1320 nm, and 1480 nm for cotton fabrics, and 1320 nm, 1420 nm, and 1600 nm for polyester fabrics).  
            \textbf{c.} Conduct an SVM model for fabric classification based on spectral characteristics, achieving a high accuracy of 98.15\% on the prediction set.
            \textbf{d.} The apple samples and experiment configuration. Apple samples with random bruises were constructed using the device shown at right.
            \textbf{e.} The acquired measurement of apples and corresponding spectral curves of bruised and normal portions. The characteristic wavelengths of apple bruises are distributed at 1060 nm, 1260 nm, and 1440 nm.
            \textbf{f.} Comparison of apple bruise detection between manual labeling (green bounding boxes) and model prediction (red bounding boxes). We employed the YOLOv5 to detect bruised portions of apples. 
            \textbf{g.} Quantitative results of apple bruise detection based on NIR and RGB images, respectively.
            }
\end{figurehere}

\section{Methods}

\subsection{Spectral modulation material preparation}

We used 16 types of organic dyes covering 400-1000 nm as spectral modulation materials for HyperspecI-V1 sensor fabrication. For the HypersepcI-V2 sensor, we utilized 10 types of organic dyes and 6 types of nano-metal oxides to cover 400-1700 nm. 
To prepare the organic dyes for photolithography processes, we mixed 0.2 g of each organic dye with 20 ml of photoresist (SU-8 2010), and utilized an ultrasonic liquid processor (NingHuai NH-1000D) at room temperature. To ensure complete dissolution and remove impurities, the mixed solution was filtered using 3 \textmu m pore size filters.  
To prepare the nano-metal oxides for photolithography processes, we used a dispersion solution (PGMEA), photoresist (SU-8 2025), and nano-metal oxide powder.  
We mixed 20 g of each material powder with 80 g $ \rm{PGMEA} $ respectively. Following a dispersion process of 48 hours using the ultrasonic liquid processor at room temperature, we obtained material dispersion fluids with a mass fraction of 20\%. To address the issue of inappropriate concentration, we mixed 10 ml of each material dispersant with 20 ml photoresist at a concentration ratio of 1:2. These mixtures were stirred for 15 minutes using the ultrasonic liquid processor. Then the filters with 3 \textmu m pore size were used to remove the impurities. 
Subsequently, we applied the spectral modulation photoresist onto the quartz substrates (JGS3), and the test smears formed at 4,000 rpm on the spin coater (Helicoater, HC220PE). We validated the spectral properties of these modulation photoresists using a spectrophotometer (PerkinElmer Lambda 950). The details of the experimental equipment, operating procedures, and result analysis are presented in Extended Data Fig. \ref{ED_Fig8_MaterialSelection},  Extended Data Fig. \ref{ED_Fig9_ModulationMaterials}, and Supplementary Section 2.

\subsection{BMSFA fabrication and sensor integration}
The fabrication of broadband multispectral filter array (BMSFA) includes a series of processes of photoresist dissolution, photomask design, substrate preparation, photoresist coating, soft bake, UV exposure, post-exposure bake, development, and hard bake (Fig. \ref{fig:1}b and Extended Data Fig. \ref{ED_Fig7_BMSFAFabrication}).
First, we prespared the photoresists with different spectral modulation properties.
Then, we dropped a solvent of one kind of photoresist onto the quartz substrate (JGS3, 4 inches), ensuring uniform distribution of the photoresist containing spectral modulation materials at 4,000 rpm on a spin coater.  
After a soft bake at 95\textdegree C for 5 minutes, we employed UV photolithography (SUSS MA6 Mask Aligner, SUSS MicroTec AG) to cure the photoresist at the designed position of the quartz substrate. 
The exposure dose of the UV lithography machine is 1,000 mJ/cm\textsuperscript{2}.
This process was conducted for different modulation materials using a designed photomask.
After a post-bake at 95\textdegree C for 10 minutes, the development removes the unexposed areas, leaving the photoresist only at the specific locations.
Then, the wafers were hard baked on a hotplate at 150°C for 5 minutes.
We repeated the above steps to pattern all 16 types of spectral modulation photoresists onto the quartz substrate.
Eventually, we poured pure SU-8 photoresist onto the finished substrate, completing the photolithography process through photoresist coating, soft bake, UV exposure, post-exposure bake, and hard baking. By following these steps, the BMSFA was prepared for our HyperspecI sensors. 

For sensor integration, the HyperspecI-V1 sensor was prepared by combining BMSFA with the Sony IMX264 chip, which covers the 400-1000 nm spectral range. The HyperspecI-V2 sensor was prepared by combining BMSFA with the Sony IMX990 chip, covering the 400-1700 nm spectral range. We utilized a laser engraving machine to remove the packaging glass from the monochrome sensor. Then, we cured the BMSFA onto the sensor surface using photoresist under UV lighting, ensuring optimal sensor integration. The details of BMSFA fabrication and sensor integration are presented in Extended Data Fig. \ref{ED_Fig7_BMSFAFabrication} and Supplementary Section 3.

\subsection{Spectral calibration}
We used a monochromator (Omno151, spectral range 200-2000 nm) to generate monochromatic light with a full width at half maximum (FWHM) of 10 nm. The monochromatic light was uniformly irradiated onto a power meter probe (Thorlabs S130VC, S132C) and the HyperspecI sensors after passing through a collimated optical path. In the automated calibration process, we developed a program to control the wavelength of monochromatic light, acquire the power meter value, and save the corresponding measurements of the HyperspecI sensors. This process was repeated for each wavelength to automatically collect the compressive sensing matrix. More details are referred to the Extended Data Fig. \ref{ED_Fig7_BMSFAFabrication}f-g and Supplementary Section 4.

\subsection{Hyperspectral image reconstruciton}
We utilized a data-driven method to reconstruct hyperspectral images (HSIs) from measurements, as shown in Extended Data Fig. \ref{ED_Fig6_SRNet}a-b.
The spectral reconstruction network (SRNet) is a hybrid neural network that combines the core features of Transformer and CNN architectures for efficient, high-precision reconstruction. It employs a Unet-shaped architecture as the baseline, whose basic component is the Spectral Attention Module (SAM) that focuses on extracting the spectral features of HSIs. SAM applies the attention mechanism in the spectral dimension rather than in the spatial dimension to reduce running time and memory cost. Besides, this strategy allows us to compute cross-covariance across spectral channels, and create attention feature maps with implicit knowledge of spectral information and global context. More details are referred to the Supplementary Section 5.2.

Our training dataset was collected using the commercial FigSpec-23 and GaiaField Pro-N17E-HR hyperspectral cameras, both integrated under a push-broom scanning mechanism, as shown in Extended Data Fig.~\ref{ED_Fig5_DatasetConstruction}. The training data comprised 96 spectral channels, with 61 channels at intervals of 10 nm in the 400-1000 nm range and 35 channels at intervals of 20 nm within the 1000-1700 nm range (see Supplementary Section 5.1 for more details).
Considering the high spatial resolution of measurements, we randomly divided the calibrated pattern into multiple sub-patterns for training, which can also avoid overfitting to a particular BMSFA encoding pattern. 
During each iteration, we randomly selected a 512 $\times$ 512 sub-pattern from the original full-resolution BMSFA pattern (2,048 $\times$ 2,048 for HyperspecI-V1 and 1,024 $\times$ 1,024 for HyperspecI-V2). 
The model was trained using the Adam optimizer ($ \beta_1 = 0.9, \beta_2 = 0.999 $) for $1 \times 10^6$ iterations. The learning rate was initialized to $ 4\times 10^{-4} $, and the Cosine Annealing scheme was adopted. We chose the Root Mean Square Error (RMSE), Mean Relative Absolute Error (MRAE), and Total Variation (TV) as the hybrid loss function. 
we trained the model on the Pytorch platform with a single NVIDIA RTX 4090 GPU. The measurements and reconstructed HSIs synthesized as RGB images are shown in Extended Data Fig. \ref{ED_Fig6_SRNet} c-d. 
The exemplar reconstructed spectra are shown in Fig. \ref{fig:2}a and Extended Data Fig. \ref{ED_Fig3_SpectralImage}. 

\subsection{Remote detection experiment}
As shown in Extended Data Fig. \ref{ED_Fig1_Moon}a, we employed a telescope (CELESTRON NEXSTAR 127SLT, 1,500 mm focal length, 127 mm aperture Maksutov-Cassegrain) to image the moon, and compared the imaging results of our HyperspecI sensor with that of line-scanning hyperspectral camera (FigSpec-23) and mosaic multispectral camera (Silios CMS-C). The target scenes include the Mare Crisium (Extended Fig. \ref{ED_Fig1_Moon}c \#2) and Mare Fecunditatis (Extended Data Fig. \ref{ED_Fig1_Moon}c \#1 and \#4) regions of the moon during the crescent phase.
Our HyperspecI's acquisition frame rate was set to 47 fps, with an exposure time of 21 ms. The mosaic multispectral camera has an acquisition frame rate of 30 fps, with an exposure time of 33 ms. The line-scanning hyperspectral camera requires approximately 100 seconds to capture one HSI frame. More details are referred to the Supplementary Section 10.

\subsection{Metamerism experiment}
Metamerism denotes that different spectra project the same color in the visible spectral range. To validate HyperspecI's capability in distinguishing materials with identical RGB values, we conducted two experiments, as shown in Extended Data Fig. \ref{ED_Fig2_Metamerism}.
First, we tested real and fake potted plants. Points with the same color were marked in Fig. \ref{ED_Fig2_Metamerism}a(ii), where the red points on the real plant and the yellow points on the fake plant have the same RGB values. Extended Data Fig. \ref{ED_Fig2_Metamerism}a(iii) shows the original measurement from our HyperspecI sensor, and the reconstructed hyperspectral image is displayed in Fig. \ref{ED_Fig2_Metamerism}a(v). We plotted the spectra of points \textbf{P1} and \textbf{P2} on both real and fake plants, as shown in Fig. \ref{ED_Fig2_Metamerism}a(iv). The spectra reveal that the leaves of the real plant exhibit distinct spectral features due to variations in chlorophyll and water content (highlighted in the blue block of Fig. \ref{ED_Fig2_Metamerism}a(iv)), and the fake plant shows completely different spectra.
Second, we tested real and fake strawberries, which present nearly identical appearances, textures, and colors, as shown in Extended Data Fig. \ref{ED_Fig2_Metamerism}b. By extracting the spectra of points \textbf{P1} and \textbf{P2} on both real and fake strawberries, we observed distinct absorption peaks at 670 nm and 750 nm in the real strawberries, whereas the spectra of the fake strawberry appeared smoother.

\subsection{Thermal stability experiment}
The experiment setup for studying the thermal stability of the BMSFA modulation mask is shown in Extended Data Fig. \ref{ED_Fig4_ThermalStability}a. This setup consists of a light source (Thorlabs SLS302 with a stabilized quartz Tungsten-Halogen lamp of 10W output optical power), an illumination module (composed of optical lens, aperture stop, field stop, beam splitter of Thorlabs VDFW5/M, and objective lens of Olympus microscope objective A 10 PL 10×0.25), a heating stage (JF-956, 30-400\textdegree C), several support components (Thorlabs CEA1400) and a fine-tune module (GCM-VC 13M). The fabricated BMSFA modulation mask was placed on the heating stage, and the temperature was sequentially increased in a step of 10\textdegree C from 20\textdegree C (room temperature) to 200\textdegree C. After the temperature stabilized at each step (waiting 10 minutes after the actual temperature reached the set temperature), we collected an image of the mask, as shown in Extended Data Fig. \ref{ED_Fig4_ThermalStability}b. More details are referred to the Supplementary Section 6.7 and the Supplementary Material Video.

Next, we placed the HyperspecI sensor on the heating stage and used it to acquire hyperspectral images of the same scene at different operation temperatures, with subsequent comparisons regarding image similarity and spectral consistency. According to the manual provided by SONY, the sensor chip's operational temperature range is from 0 to 50\textdegree C, with the common operating surface temperature being around 37\textdegree C at room temperature (20 \textdegree C). To assess the sensor's reconstruction performance at varying temperatures, the heating stage was incrementally heated from 40\textdegree C to 70\textdegree C with 10\textdegree C intervals. At each temperature, the sensor was powered on for 1 hour to achieve temperature stabilization. Extended Data Fig. \ref{ED_Fig4_ThermalStability}d-f illustrates the hyperspectral imaging performance of the same scene at different temperatures.

\newpage

\noindent\textbf{Data availability} All data generated or analyzed during this study are included in this published article and the public repository at the following link: https://github.com/bianlab/Hyperspectral-imaging-dataset.\\

\noindent\textbf{Code availability} The demo code of this work is available from the public repository at the following link: https://github.com/bianlab/HyperspecI.\\

\noindent\textbf{Acknowledgements} This work was supported by the National Natural Science Foundation of China (62322502, 61827901, 62088101, 61971045).\\

\noindent\textbf{Author contributions} L.B., Z.W. and Y.Z.Z. conceived the idea. Z.W., Y.Z.Z., C.Y., and W.F. conducted material optical performance tests and photoresist preparation. Z.W. and Y.Z.Z. designed and fabricated the optical filter arrays. Y.Z.Z. and Y.N.Z designed and implemented sensor integration. Z.W. and Y.Z.Z. developed the reconstruction algorithms and conducted model training. Z.W. and Y.Z.Z. calibrated the sensors and tested their imaging performance. L.L., X.P., Y.N.Z., and J.J.Z designed and implemented the experiments of chlorophyll detection, SSC detection, textile classification, and apple bruise detection.
Y.Z.Z., Q.M., and Y.N.Z conducted blood oxygen and water quality monitoring experiments. L.B., Z.W., Y.Z.Z., Y.N.Z., C.Y., L.L., C.Z., and J.Z. prepared the figures and wrote the manuscript with input from all the authors. L.B. and J.Z supervised the project.\\

\noindent\textbf{Competing interests}
L.B., Z.W., Y.Z.Z, and J.Z. hold patents on technologies related to the devices developed in this work (China patent numbers ZL 2022 1 0764166.5, ZL 2022 1 0764143.4, ZL 2022 1 0764141.5, ZL 2019 1 0441784.4, ZL 2019 1 0482098.1, and ZL 2019 1 1234638.0) and submitted related patent applications.\\

{\color{blue}
\noindent\textbf{Additional information}\\
\noindent\textbf{Supplementary information}
Supplementary Information is available for this paper at xxx.\\
\noindent\textbf{Correspondence and requests for materials} should be addressed to Liheng Bian.\\
\noindent\textbf{Peer review information} Nature thanks the anonymous reviewers for their contribution to the peer review of this work.\\
\noindent\textbf{Reprints and permissions information} is available at http://www. nature.com/reprints.
}

\newpage

\setcounter{figure}{0}
\captionsetup[figure]{name=Extended Data Fig.}

\renewcommand{\figurename}{Extended Data Fig.} 
\captionsetup{font={small}}
\begin{figurehere}
	\centering
	\caption{\label{ED_Fig1_Moon}
            {\textbf{Dynamic remote detection experiment in low-light environment (a cloudy and foggy night) with imaging comparison among our HyperspecI sensor, line-scanning hyperspectral camera (FigSpec-23), and mosaic multispectral camera (Silios CMS-C).}} 
            {\textbf{a.}} The experiment configuration. A telescope (CELESTRON NEXSTAR 127SLT, 1,500 mm focal length, 127 mm aperture Maksutov-Cassegrain) was employed to image the moon combined with different cameras.
            {\textbf{b.}} The lunar spectrum comparison.
            {\textbf{c.}} The HSI results by different cameras. The results of the HyperspecI sensor present fine details of lunar topography, and the reconstructed spectrum corresponds well with the ground truth. In contrast, the results of the mosaic multispectral camera contain serious measurement noise due to limited light throughput, and the topography details are buried. The results of the line-scanning hyperspectral camera suffer from a similar degradation, and severe scanning overlapping exists since the moon was moving during the line-scanning process. The above experiment demonstrates the unique high-light throughput advantage of our HyperspecI sensor, which leads to high imaging SNR that enables the acquisition of dynamic, remote, and fine details in low-light conditions.
            {\textbf{d.}} The dynamic imaging results by different cameras.
            }   
\end{figurehere}

\begin{figurehere}
	\centering
	\caption{\label{ED_Fig2_Metamerism}
            {\textbf{Metamerism experiment.}}
            {\textbf{a.}} Hyperspectral imaging results of our HyperspecI sensor on real and fake potted plants with the same color but different spectra. (i) RGB images of real and potted fake plants. (ii) Locations of real and fake plants of the same color are marked with red points and yellow points, respectively. (iii) The raw measurement of the HyperspecI sensor. (iv) Reconstructed spectra of metamerism locations. (v) Synthesized RGB image of the reconstructed HSI.
            {\textbf{b.}} Hyperspectral imaging results of our HyperspecI sensor on real and fake strawberries with the same color but different spectra. (i) RGB images of real and fake strawberries. (ii) Locations of real and fake strawberries of the same color are marked with red points and yellow points, respectively. (iii) The raw measurement of the HyperspecI sensor.(iv) Reconstructed spectra of metamerism locations. (v) Synthesized RGB image of the reconstructed HSI.
            }   
\end{figurehere}

\begin{figure}[!htbp]
	\centering
	\caption{\label{ED_Fig3_SpectralImage}
            {\textbf{Exemplar HSI results by our HyperspecI sensors.}} {\textbf{a-d.}} HSI results of four different indoor and outdoor scenes. The measurements were acquired by our HyperspecI sensors. The hyperspectral images were reconstructed via SRNet. Synthesized RGB images and several spectral images are presented. The spectral comparison between reconstructed spectra (RS) and ground truth (GT, acquired by the commercial spectrometers of Ocean Optics USB 2000+ and NIR-Quest 512) are also represented (denoted by solid and dashed lines, respectively).
            }   
\end{figure}

\clearpage
\begin{figure}[!htbp]
    \centering
     \captionsetup{font={small}}
    \caption{\label{ED_Fig4_ThermalStability}
            {\textbf{Thermal stability test of the BMSFA modulation mask and HyperspecI sensor.}} 
            {\textbf{a.}} The experimental configuration for BMSFA thermal stability test, comprising an optical system (including components of light source, illuminating system, camera, camera tube, beam splitter, objective lens, etc.) for uniform light illumination on the target, mechanical elements (featuring a manual focusing module, heating stage, translation stage, main support, etc.) for precise control of the target’s observation position and target heating, and the modulation mask.
            {\textbf{b.}} The visual representations of the modulation mask at different temperatures. These observations reveal that the modulation mask is stable under different temperature conditions, maintaining its structural integrity and properties.
            {\textbf{c.}} The experiment configuration for sensor thermal stability test. The HyperspecI sensor was fixed on a heating stage with controllable temperatures ranging from 40 \textdegree C to 70 \textdegree C. Measurements were acquired after each thermal step reached stability, with the sensor operating for 1 hour at each temperature.
            {\textbf{d.}} Similarity evaluation results of raw data. The SSIM and PSNR measurements consistently indicate that the camera’s performance remains unaffected across different operating temperatures.
            {\textbf{e.}} The acquired raw data and corresponding HSI reconstruction results at different temperatures.
            {\textbf{f.}} Reconstructed spectral comparison of different regions. We calculated the Pearson correlation coefficients of the spectra in the same region at different temperatures. The minimum correlation coefficient for each region is 0.99, indicating that the sensor’s spectral reconstruction performance is robust to temperature variations.
            }   
\end{figure}

\begin{figure}[!htbp]
	\centering
	\caption{\label{ED_Fig5_DatasetConstruction}
            {\textbf{Hyperspectral image dataset construction.}} {\textbf{a.}} The system to collect hyperspectral image dataset. Our dataset was mainly captured using the commercial FigSpec-23 (400-1000 nm $@$ 960×1,230 pixels, 2.5 nm interval) and GaiaField Pro-N17E-HR (900-1700 nm, $@$ 640×666 pixels, 5 nm interval) hyperspectral cameras, both integrated under a push-broom scanning mechanism. Measurements were acquired using our HyperspecI (V1 for 400-1000 nm and V2 for 400-1700 nm).
            {\textbf{b.}} Image registration between the two commercial hyperspectral cameras. The scale-invariant feature transform (SIFT) technique was employed to align the field of view.
            {\textbf{c.}} The visualization of our constructed hyperspectral image dataset. After data registration, there yields the hyperspectral image dataset comprising 1,000 scenes (500 outdoor scenes and 500 indoor scenes), covering the entire spectral range of 400-1700 nm, with a spatial resolution of 640 × 666 pixels and a total number of 131 spectral bands at 10 nm intervals.
            }   
\end{figure}

\begin{figure}[!htbp]
	\centering
	\caption{\label{ED_Fig6_SRNet}
            {\textbf{The spectral reconstruction network (SRNet) structure and exemplar reconstructed results.}}
            {\textbf{a.}} The overall framework of SRNet. SRNet is a hybrid neural network that combines the core features of Transformer and CNN architectures for efficient, high-precision reconstruction.
            {\textbf{b.}} The framework of the Spectral Attention Module (SAM). SAM is the basic component of SRNet, which calculates the attention across spectral channel dimensions, extracting the spectral features of HSIs. 
            {\textbf{c-d.}} Measurements and corresponding spectral reconstruction result (presented as the synthesized RGB form). The measurements were acquired using the HyperspecI-V1 sensor. Close-ups are provided, marked in the measurements with rectangular outlines.
            }   
\end{figure}

\begin{figure}[!htbp]
	\centering
	\caption{\label{ED_Fig7_BMSFAFabrication}
            {\textbf{The preparation, integration, and calibration demonstration of HyperspecI sensors.}} {\textbf{a.}} The demonstration of BMSFA photolithography fabrication.
            {\textbf{b.}} Display of integrated HyperspecI-V1 sensor.
            {\textbf{c.}} Display of integrated and packaged HyperspecI-V2 sensor.
            {\textbf{d.}} Photolithography mask used for BMSF fabrication. Multiple lithography operations can be achieved using this single mask.
            {\textbf{e.}} BMSFA fabrication and its microstructure, including microscopic images during the fabrication process.
            {\textbf{f.}} Display of the HyperspecI-V1 sensor's sensing matrix calibrated with monochromatic light in several spectral bands (550 nm, 650 nm, 750 nm).
            {\textbf{g.}} Display of the HyperspecI-V2 sensor's sensing matrix calibrated with monochromatic light in several spectral bands (600 nm, 800 nm, 1300 nm).
            }   
\end{figure}

\clearpage

\begin{figurehere}
	\centering
	\caption{\label{ED_Fig8_MaterialSelection}
            {\textbf{The material selection and BMSFA design study.}} 
            {\textbf{a.}} The evolutionary optimization based material selection method for BMSFA design. This method starts with an initially selected subset of materials and iterates through the operations, including survival of the fittest, crossover, mutation, and random replacement. The iterative process ends when it converges to the optimal accuracy performance on the hyperspectral image dataset.
            {\textbf{b.}} The preprocess and analysis of the massive hyperspectral image data through the dimensionality reduction technique. We analyzed the distribution of the hyperspectral images using the PCA method. We calculated the information loss (reconstruction error) in different latent dimensions and compressed ratios to determine the potential compressive dimension. We can see that the reconstruction error is low with the dimension number being ten at 400-1000 nm range and six at 1000-1700 nm range, which demonstrates the sparsity of HSI in the spectral dimension.
            {\textbf{c.}} The spectral fidelity under different numbers of modulation filters selected by the material selection method. The signal-to-noise ratio of input measurements was set as 20 dB. It further validates the reasonability of the number and selection of broadband filters, and shows that the current choice of our HyperspecI sensor prototypes is optimal considering the tradeoff between spectral and spatial resolution.
            {\textbf{d.}} The organic dyes and nano-metal oxides prepared for BMSFA design.
            {\textbf{e.}} The correlation coefficient map of the prepared 35 materials.
            }   
\end{figurehere}

\begin{figure}[ht]
	\centering
	\caption{\label{ED_Fig9_ModulationMaterials}
            {\textbf{Modulation material preparation and transmission spectra measurements.}} {\textbf{a.}} Schematic diagram depicting the production of experimental smears using spectral modulation materials. This process follows the steps of weighting, mixing, filtering, and spin coating.
            {\textbf{b.}} Schematic diagram of the optical path for transmission spectra measurements of spectral modulation materials. 
            {\textbf{c.}} The smears of organic dyes, employing photoresist as a carrier, are obtained through spin coating.
            {\textbf{d.}} The transmission spectra of organic dyes.
            {\textbf{e.}} The smears of nano-metal oxides, utilizing photoresist and dispersant as carriers, are obtained through spin coating.
            {\textbf{f.}} The transmission spectra of nano-metal oxides at the optimum concentration. 
            }   
\end{figure}

\newpage
\setcounter{table}{0}
\captionsetup[table]{name=Extended Data Table.}
\begin{table}[!ht]
    \caption{Comparison of different snapshot hyperspectral imaging techniques.}
    \label{Extended_Table_Comparison}
    \begin{center}
    \end{center}
\end{table}


\begin{thebibliography}{49}
\ifx \bisbn   \undefined \def \bisbn  #1{ISBN #1}\fi
\ifx \binits  \undefined \def \binits#1{#1}\fi
\ifx \bauthor  \undefined \def \bauthor#1{#1}\fi
\ifx \batitle  \undefined \def \batitle#1{#1}\fi
\ifx \bjtitle  \undefined \def \bjtitle#1{#1}\fi
\ifx \bvolume  \undefined \def \bvolume#1{\textbf{#1}}\fi
\ifx \byear  \undefined \def \byear#1{#1}\fi
\ifx \bissue  \undefined \def \bissue#1{#1}\fi
\ifx \bfpage  \undefined \def \bfpage#1{#1}\fi
\ifx \blpage  \undefined \def \blpage #1{#1}\fi
\ifx \burl  \undefined \def \burl#1{\textsf{#1}}\fi
\ifx \doiurl  \undefined \def \doiurl#1{\url{https://doi.org/#1}}\fi
\ifx \betal  \undefined \def \betal{\textit{et al.}}\fi
\ifx \binstitute  \undefined \def \binstitute#1{#1}\fi
\ifx \binstitutionaled  \undefined \def \binstitutionaled#1{#1}\fi
\ifx \bctitle  \undefined \def \bctitle#1{#1}\fi
\ifx \beditor  \undefined \def \beditor#1{#1}\fi
\ifx \bpublisher  \undefined \def \bpublisher#1{#1}\fi
\ifx \bbtitle  \undefined \def \bbtitle#1{#1}\fi
\ifx \bedition  \undefined \def \bedition#1{#1}\fi
\ifx \bseriesno  \undefined \def \bseriesno#1{#1}\fi
\ifx \blocation  \undefined \def \blocation#1{#1}\fi
\ifx \bsertitle  \undefined \def \bsertitle#1{#1}\fi
\ifx \bsnm \undefined \def \bsnm#1{#1}\fi
\ifx \bsuffix \undefined \def \bsuffix#1{#1}\fi
\ifx \bparticle \undefined \def \bparticle#1{#1}\fi
\ifx \barticle \undefined \def \barticle#1{#1}\fi
\bibcommenthead
\ifx \bconfdate \undefined \def \bconfdate #1{#1}\fi
\ifx \botherref \undefined \def \botherref #1{#1}\fi
\ifx \url \undefined \def \url#1{\textsf{#1}}\fi
\ifx \bchapter \undefined \def \bchapter#1{#1}\fi
\ifx \bbook \undefined \def \bbook#1{#1}\fi
\ifx \bcomment \undefined \def \bcomment#1{#1}\fi
\ifx \oauthor \undefined \def \oauthor#1{#1}\fi
\ifx \citeauthoryear \undefined \def \citeauthoryear#1{#1}\fi
\ifx \endbibitem  \undefined \def \endbibitem {}\fi
\ifx \bconflocation  \undefined \def \bconflocation#1{#1}\fi
\ifx \arxivurl  \undefined \def \arxivurl#1{\textsf{#1}}\fi
\csname PreBibitemsHook\endcsname

\bibitem{landgrebe2002hyperspectral}
\begin{barticle}
\bauthor{\bsnm{Landgrebe}, \binits{D.}}:
\batitle{Hyperspectral image data analysis}.
\bjtitle{IEEE Signal Proc. Mag.}
\bvolume{19}(\bissue{1}),
\bfpage{17}--\blpage{28}
(\byear{2002})
\end{barticle}
\endbibitem

\bibitem{li2019deep}
\begin{barticle}
\bauthor{\bsnm{Li}, \binits{S.}},
\bauthor{\bsnm{Song}, \binits{W.}},
\bauthor{\bsnm{Fang}, \binits{L.}},
\bauthor{\bsnm{Chen}, \binits{Y.}},
\bauthor{\bsnm{Ghamisi}, \binits{P.}},
\bauthor{\bsnm{Benediktsson}, \binits{J.A.}}:
\batitle{Deep learning for hyperspectral image classification: {A}n overview}.
\bjtitle{IEEE T. Geosci. Remote}
\bvolume{57}(\bissue{9}),
\bfpage{6690}--\blpage{6709}
(\byear{2019})
\end{barticle}
\endbibitem

\bibitem{backman2000detection}
\begin{barticle}
\bauthor{\bsnm{Backman}, \binits{V.}},
\bauthor{\bsnm{Wallace}, \binits{M.B.}},
\bauthor{\bsnm{Perelman}, \binits{L.}},
\bauthor{\bsnm{Arendt}, \binits{J.}},
\bauthor{\bsnm{Gurjar}, \binits{R.}},
\bauthor{\bsnm{M{\"u}ller}, \binits{M.}},
\bauthor{\bsnm{Zhang}, \binits{Q.}},
\bauthor{\bsnm{Zonios}, \binits{G.}},
\bauthor{\bsnm{Kline}, \binits{E.}},
\bauthor{\bsnm{McGillican}, \binits{T.}}, \betal:
\batitle{Detection of preinvasive cancer cells}.
\bjtitle{Nature}
\bvolume{406}(\bissue{6791}),
\bfpage{35}--\blpage{36}
(\byear{2000})
\end{barticle}
\endbibitem

\bibitem{hadoux2019non}
\begin{barticle}
\bauthor{\bsnm{Hadoux}, \binits{X.}},
\bauthor{\bsnm{Hui}, \binits{F.}},
\bauthor{\bsnm{Lim}, \binits{J.K.}},
\bauthor{\bsnm{Masters}, \binits{C.L.}},
\bauthor{\bsnm{P{\'e}bay}, \binits{A.}},
\bauthor{\bsnm{Chevalier}, \binits{S.}},
\bauthor{\bsnm{Ha}, \binits{J.}},
\bauthor{\bsnm{Loi}, \binits{S.}},
\bauthor{\bsnm{Fowler}, \binits{C.J.}},
\bauthor{\bsnm{Rowe}, \binits{C.}}, \betal:
\batitle{Non-invasive in vivo hyperspectral imaging of the retina for potential biomarker use in {A}lzheimer’s disease}.
\bjtitle{Nat. Commun.}
\bvolume{10}(\bissue{1}),
\bfpage{4227}
(\byear{2019})
\end{barticle}
\endbibitem

\bibitem{mehl2004development}
\begin{barticle}
\bauthor{\bsnm{Mehl}, \binits{P.M.}},
\bauthor{\bsnm{Chen}, \binits{Y.-R.}},
\bauthor{\bsnm{Kim}, \binits{M.S.}},
\bauthor{\bsnm{Chan}, \binits{D.E.}}:
\batitle{Development of hyperspectral imaging technique for the detection of apple surface defects and contaminations}.
\bjtitle{J. Food Eng.}
\bvolume{61}(\bissue{1}),
\bfpage{67}--\blpage{81}
(\byear{2004})
\end{barticle}
\endbibitem

\bibitem{yang2019single}
\begin{barticle}
\bauthor{\bsnm{Yang}, \binits{Z.}},
\bauthor{\bsnm{Albrow-Owen}, \binits{T.}},
\bauthor{\bsnm{Cui}, \binits{H.}},
\bauthor{\bsnm{Alexander-Webber}, \binits{J.}},
\bauthor{\bsnm{Gu}, \binits{F.}},
\bauthor{\bsnm{Wang}, \binits{X.}},
\bauthor{\bsnm{Wu}, \binits{T.-C.}},
\bauthor{\bsnm{Zhuge}, \binits{M.}},
\bauthor{\bsnm{Williams}, \binits{C.}},
\bauthor{\bsnm{Wang}, \binits{P.}}, \betal:
\batitle{Single-nanowire spectrometers}.
\bjtitle{Science}
\bvolume{365}(\bissue{6457}),
\bfpage{1017}--\blpage{1020}
(\byear{2019})
\end{barticle}
\endbibitem

\bibitem{green1998imaging}
\begin{barticle}
\bauthor{\bsnm{Green}, \binits{R.O.}},
\bauthor{\bsnm{Eastwood}, \binits{M.L.}},
\bauthor{\bsnm{Sarture}, \binits{C.M.}},
\bauthor{\bsnm{Chrien}, \binits{T.G.}},
\bauthor{\bsnm{Aronsson}, \binits{M.}},
\bauthor{\bsnm{Chippendale}, \binits{B.J.}},
\bauthor{\bsnm{Faust}, \binits{J.A.}},
\bauthor{\bsnm{Pavri}, \binits{B.E.}},
\bauthor{\bsnm{Chovit}, \binits{C.J.}},
\bauthor{\bsnm{Solis}, \binits{M.}}, \betal:
\batitle{Imaging spectroscopy and the airborne visible/infrared imaging spectrometer ({AVIRIS})}.
\bjtitle{Remote Sens. Environ.}
\bvolume{65}(\bissue{3}),
\bfpage{227}--\blpage{248}
(\byear{1998})
\end{barticle}
\endbibitem

\bibitem{pian2017compressive}
\begin{barticle}
\bauthor{\bsnm{Pian}, \binits{Q.}},
\bauthor{\bsnm{Yao}, \binits{R.}},
\bauthor{\bsnm{Sinsuebphon}, \binits{N.}},
\bauthor{\bsnm{Intes}, \binits{X.}}:
\batitle{Compressive hyperspectral time-resolved wide-field fluorescence lifetime imaging}.
\bjtitle{Nat. Photonics}
\bvolume{11}(\bissue{7}),
\bfpage{411}--\blpage{414}
(\byear{2017})
\end{barticle}
\endbibitem

\bibitem{descour1995computed}
\begin{barticle}
\bauthor{\bsnm{Descour}, \binits{M.}},
\bauthor{\bsnm{Dereniak}, \binits{E.}}:
\batitle{Computed-tomography imaging spectrometer: experimental calibration and reconstruction results}.
\bjtitle{Appl. Opt.}
\bvolume{34}(\bissue{22}),
\bfpage{4817}--\blpage{4826}
(\byear{1995})
\end{barticle}
\endbibitem

\bibitem{wagadarikar2008single}
\begin{barticle}
\bauthor{\bsnm{Wagadarikar}, \binits{A.}},
\bauthor{\bsnm{John}, \binits{R.}},
\bauthor{\bsnm{Willett}, \binits{R.}},
\bauthor{\bsnm{Brady}, \binits{D.}}:
\batitle{Single disperser design for coded aperture snapshot spectral imaging}.
\bjtitle{Appl. Opt.}
\bvolume{47}(\bissue{10}),
\bfpage{44}--\blpage{51}
(\byear{2008})
\end{barticle}
\endbibitem

\bibitem{arguello2014colored}
\begin{barticle}
\bauthor{\bsnm{Arguello}, \binits{H.}},
\bauthor{\bsnm{Arce}, \binits{G.R.}}:
\batitle{Colored coded aperture design by concentration of measure in compressive spectral imaging}.
\bjtitle{IEEE T. Image Process.}
\bvolume{23}(\bissue{4}),
\bfpage{1896}--\blpage{1908}
(\byear{2014})
\end{barticle}
\endbibitem

\bibitem{geelen2014compact}
\begin{bchapter}
\bauthor{\bsnm{Geelen}, \binits{B.}},
\bauthor{\bsnm{Tack}, \binits{N.}},
\bauthor{\bsnm{Lambrechts}, \binits{A.}}:
\bctitle{A compact snapshot multispectral imager with a monolithically integrated per-pixel filter mosaic}.
In: \bbtitle{Advanced Fabrication Technologies for Micro/nano Optics and Photonics VII},
vol. \bseriesno{8974},
pp. \bfpage{80}--\blpage{87}
(\byear{2014})
\end{bchapter}
\endbibitem

\bibitem{yesilkoy2019ultrasensitive}
\begin{barticle}
\bauthor{\bsnm{Yesilkoy}, \binits{F.}},
\bauthor{\bsnm{Arvelo}, \binits{E.R.}},
\bauthor{\bsnm{Jahani}, \binits{Y.}},
\bauthor{\bsnm{Liu}, \binits{M.}},
\bauthor{\bsnm{Tittl}, \binits{A.}},
\bauthor{\bsnm{Cevher}, \binits{V.}},
\bauthor{\bsnm{Kivshar}, \binits{Y.}},
\bauthor{\bsnm{Altug}, \binits{H.}}:
\batitle{Ultrasensitive hyperspectral imaging and biodetection enabled by dielectric metasurfaces}.
\bjtitle{Nat. Photonics}
\bvolume{13}(\bissue{6}),
\bfpage{390}--\blpage{396}
(\byear{2019})
\end{barticle}
\endbibitem

\bibitem{faraji2019hyperspectral}
\begin{barticle}
\bauthor{\bsnm{Faraji-Dana}, \binits{M.}},
\bauthor{\bsnm{Arbabi}, \binits{E.}},
\bauthor{\bsnm{Kwon}, \binits{H.}},
\bauthor{\bsnm{Kamali}, \binits{S.M.}},
\bauthor{\bsnm{Arbabi}, \binits{A.}},
\bauthor{\bsnm{Bartholomew}, \binits{J.G.}},
\bauthor{\bsnm{Faraon}, \binits{A.}}:
\batitle{Hyperspectral imager with folded metasurface optics}.
\bjtitle{ACS Photonics}
\bvolume{6}(\bissue{8}),
\bfpage{2161}--\blpage{2167}
(\byear{2019})
\end{barticle}
\endbibitem

\bibitem{xiong2022dynamic}
\begin{barticle}
\bauthor{\bsnm{Xiong}, \binits{J.}},
\bauthor{\bsnm{Cai}, \binits{X.}},
\bauthor{\bsnm{Cui}, \binits{K.}},
\bauthor{\bsnm{Huang}, \binits{Y.}},
\bauthor{\bsnm{Yang}, \binits{J.}},
\bauthor{\bsnm{Zhu}, \binits{H.}},
\bauthor{\bsnm{Li}, \binits{W.}},
\bauthor{\bsnm{Hong}, \binits{B.}},
\bauthor{\bsnm{Rao}, \binits{S.}},
\bauthor{\bsnm{Zheng}, \binits{Z.}}, \betal:
\batitle{Dynamic brain spectrum acquired by a real-time ultraspectral imaging chip with reconfigurable metasurfaces}.
\bjtitle{Optica}
\bvolume{9}(\bissue{5}),
\bfpage{461}--\blpage{468}
(\byear{2022})
\end{barticle}
\endbibitem

\bibitem{he2024meta}
\begin{botherref}
\oauthor{\bsnm{He}, \binits{H.}},
\oauthor{\bsnm{Zhang}, \binits{Y.}},
\oauthor{\bsnm{Shao}, \binits{Y.}},
\oauthor{\bsnm{Zhang}, \binits{Y.}},
\oauthor{\bsnm{Geng}, \binits{G.}},
\oauthor{\bsnm{Li}, \binits{J.}},
\oauthor{\bsnm{Li}, \binits{X.}},
\oauthor{\bsnm{Wang}, \binits{Y.}},
\oauthor{\bsnm{Bian}, \binits{L.}},
\oauthor{\bsnm{Zhang}, \binits{J.}}, et al.:
Meta-attention network based spectral reconstruction with snapshot near-infrared metasurface.
Adv. Mater.,
2313357
(2024)
\end{botherref}
\endbibitem

\bibitem{wang2019single}
\begin{barticle}
\bauthor{\bsnm{Wang}, \binits{Z.}},
\bauthor{\bsnm{Yi}, \binits{S.}},
\bauthor{\bsnm{Chen}, \binits{A.}},
\bauthor{\bsnm{Zhou}, \binits{M.}},
\bauthor{\bsnm{Luk}, \binits{T.S.}},
\bauthor{\bsnm{James}, \binits{A.}},
\bauthor{\bsnm{Nogan}, \binits{J.}},
\bauthor{\bsnm{Ross}, \binits{W.}},
\bauthor{\bsnm{Joe}, \binits{G.}},
\bauthor{\bsnm{Shahsafi}, \binits{A.}}, \betal:
\batitle{Single-shot on-chip spectral sensors based on photonic crystal slabs}.
\bjtitle{Nat. Commun.}
\bvolume{10}(\bissue{1}),
\bfpage{1020}
(\byear{2019})
\end{barticle}
\endbibitem

\bibitem{yako2023video}
\begin{barticle}
\bauthor{\bsnm{Yako}, \binits{M.}},
\bauthor{\bsnm{Yamaoka}, \binits{Y.}},
\bauthor{\bsnm{Kiyohara}, \binits{T.}},
\bauthor{\bsnm{Hosokawa}, \binits{C.}},
\bauthor{\bsnm{Noda}, \binits{A.}},
\bauthor{\bsnm{Tack}, \binits{K.}},
\bauthor{\bsnm{Spooren}, \binits{N.}},
\bauthor{\bsnm{Hirasawa}, \binits{T.}},
\bauthor{\bsnm{Ishikawa}, \binits{A.}}:
\batitle{Video-rate hyperspectral camera based on a {CMOS}-compatible random array of {Fabry}--{P}{\'e}rot filters}.
\bjtitle{Nat. Photonics}
\bvolume{17}(\bissue{3}),
\bfpage{218}--\blpage{223}
(\byear{2023})
\end{barticle}
\endbibitem

\bibitem{kim2023aperture}
\begin{botherref}
\oauthor{\bsnm{Kim}, \binits{T.}},
\oauthor{\bsnm{Lee}, \binits{K.C.}},
\oauthor{\bsnm{Baek}, \binits{N.}},
\oauthor{\bsnm{Chae}, \binits{H.}},
\oauthor{\bsnm{Lee}, \binits{S.A.}}:
Aperture-encoded snapshot hyperspectral imaging with a lensless camera.
APL Photonics
\textbf{8}(6)
(2023)
\end{botherref}
\endbibitem

\bibitem{redding2013compact}
\begin{barticle}
\bauthor{\bsnm{Redding}, \binits{B.}},
\bauthor{\bsnm{Liew}, \binits{S.F.}},
\bauthor{\bsnm{Sarma}, \binits{R.}},
\bauthor{\bsnm{Cao}, \binits{H.}}:
\batitle{Compact spectrometer based on a disordered photonic chip}.
\bjtitle{Nat. Photonics}
\bvolume{7}(\bissue{9}),
\bfpage{746}--\blpage{751}
(\byear{2013})
\end{barticle}
\endbibitem

\bibitem{monakhova2020spectral}
\begin{barticle}
\bauthor{\bsnm{Monakhova}, \binits{K.}},
\bauthor{\bsnm{Yanny}, \binits{K.}},
\bauthor{\bsnm{Aggarwal}, \binits{N.}},
\bauthor{\bsnm{Waller}, \binits{L.}}:
\batitle{Spectral diffusercam: lensless snapshot hyperspectral imaging with a spectral filter array}.
\bjtitle{Optica}
\bvolume{7}(\bissue{10}),
\bfpage{1298}--\blpage{1307}
(\byear{2020})
\end{barticle}
\endbibitem

\bibitem{jeon2019compact}
\begin{barticle}
\bauthor{\bsnm{Jeon}, \binits{D.S.}},
\bauthor{\bsnm{Baek}, \binits{S.-H.}},
\bauthor{\bsnm{Yi}, \binits{S.}},
\bauthor{\bsnm{Fu}, \binits{Q.}},
\bauthor{\bsnm{Dun}, \binits{X.}},
\bauthor{\bsnm{Heidrich}, \binits{W.}},
\bauthor{\bsnm{Kim}, \binits{M.H.}}:
\batitle{Compact snapshot hyperspectral imaging with diffracted rotation}.
\bjtitle{ACM Trans. Graph.}
\bvolume{38}(\bissue{4}),
\bfpage{13}
(\byear{2019})
\end{barticle}
\endbibitem

\bibitem{cortes2019monitoring}
\begin{barticle}
\bauthor{\bsnm{Cort{\'e}s}, \binits{V.}},
\bauthor{\bsnm{Blasco}, \binits{J.}},
\bauthor{\bsnm{Aleixos}, \binits{N.}},
\bauthor{\bsnm{Cubero}, \binits{S.}},
\bauthor{\bsnm{Talens}, \binits{P.}}:
\batitle{Monitoring strategies for quality control of agricultural products using visible and near-infrared spectroscopy: {A} review}.
\bjtitle{Trends Food Sci. Technol.}
\bvolume{85},
\bfpage{138}--\blpage{148}
(\byear{2019})
\end{barticle}
\endbibitem

\bibitem{limantara2015analysis}
\begin{barticle}
\bauthor{\bsnm{Limantara}, \binits{L.}},
\bauthor{\bsnm{Dettling}, \binits{M.}},
\bauthor{\bsnm{Indrawati}, \binits{R.}},
\bauthor{\bsnm{Brotosudarmo}, \binits{T.H.P.}}, \betal:
\batitle{Analysis on the chlorophyll content of commercial green leafy vegetables}.
\bjtitle{Procedia Chem.}
\bvolume{14},
\bfpage{225}--\blpage{231}
(\byear{2015})
\end{barticle}
\endbibitem

\bibitem{li2022calibration}
\begin{barticle}
\bauthor{\bsnm{Li}, \binits{L.}},
\bauthor{\bsnm{Huang}, \binits{W.}},
\bauthor{\bsnm{Wang}, \binits{Z.}},
\bauthor{\bsnm{Liu}, \binits{S.}},
\bauthor{\bsnm{He}, \binits{X.}},
\bauthor{\bsnm{Fan}, \binits{S.}}:
\batitle{Calibration transfer between developed portable {Vis/NIR} devices for detection of soluble solids contents in apple}.
\bjtitle{Postharvest Biol. Technol.}
\bvolume{183},
\bfpage{111720}
(\byear{2022})
\end{barticle}
\endbibitem

\bibitem{ma2021rapid}
\begin{barticle}
\bauthor{\bsnm{Ma}, \binits{T.}},
\bauthor{\bsnm{Xia}, \binits{Y.}},
\bauthor{\bsnm{Inagaki}, \binits{T.}},
\bauthor{\bsnm{Tsuchikawa}, \binits{S.}}:
\batitle{Rapid and nondestructive evaluation of soluble solids content {(SSC)} and firmness in apple using {Vis}--{NIR} spatially resolved spectroscopy}.
\bjtitle{Postharvest Biol.Technol.}
\bvolume{173},
\bfpage{111417}
(\byear{2021})
\end{barticle}
\endbibitem

\bibitem{liu2020qualitative}
\begin{barticle}
\bauthor{\bsnm{Liu}, \binits{Z.}},
\bauthor{\bsnm{Li}, \binits{W.}},
\bauthor{\bsnm{Wei}, \binits{Z.}}:
\batitle{Qualitative classification of waste textiles based on near infrared spectroscopy and the convolutional network}.
\bjtitle{Text. Res. J.}
\bvolume{90}(\bissue{9-10}),
\bfpage{1057}--\blpage{1066}
(\byear{2020})
\end{barticle}
\endbibitem

\bibitem{kim2014all}
\begin{barticle}
\bauthor{\bsnm{Kim}, \binits{S.}},
\bauthor{\bsnm{Marelli}, \binits{B.}},
\bauthor{\bsnm{Brenckle}, \binits{M.A.}},
\bauthor{\bsnm{Mitropoulos}, \binits{A.N.}},
\bauthor{\bsnm{Gil}, \binits{E.-S.}},
\bauthor{\bsnm{Tsioris}, \binits{K.}},
\bauthor{\bsnm{Tao}, \binits{H.}},
\bauthor{\bsnm{Kaplan}, \binits{D.L.}},
\bauthor{\bsnm{Omenetto}, \binits{F.G.}}:
\batitle{All-water-based electron-beam lithography using silk as a resist}.
\bjtitle{Nat. Nanotechnol.}
\bvolume{9}(\bissue{4}),
\bfpage{306}--\blpage{310}
(\byear{2014})
\end{barticle}
\endbibitem

\bibitem{yu20172d}
\begin{barticle}
\bauthor{\bsnm{Yu}, \binits{S.}},
\bauthor{\bsnm{Wu}, \binits{X.}},
\bauthor{\bsnm{Wang}, \binits{Y.}},
\bauthor{\bsnm{Guo}, \binits{X.}},
\bauthor{\bsnm{Tong}, \binits{L.}}:
\batitle{{2D materials for optical modulation: challenges and opportunities}}.
\bjtitle{Adv. Mater.}
\bvolume{29}(\bissue{14}),
\bfpage{1606128}
(\byear{2017})
\end{barticle}
\endbibitem

\bibitem{zheng2015illumination}
\begin{bchapter}
\bauthor{\bsnm{Zheng}, \binits{Y.}},
\bauthor{\bsnm{Sato}, \binits{I.}},
\bauthor{\bsnm{Sato}, \binits{Y.}}:
\bctitle{Illumination and reflectance spectra separation of a hyperspectral image meets low-rank matrix factorization}.
In: \bbtitle{Proc. IEEE Conf. Comput. Vis. Pattern Recog.},
pp. \bfpage{1779}--\blpage{1787}
(\byear{2015})
\end{bchapter}
\endbibitem

\bibitem{abdar2021review}
\begin{barticle}
\bauthor{\bsnm{Abdar}, \binits{M.}},
\bauthor{\bsnm{Pourpanah}, \binits{F.}},
\bauthor{\bsnm{Hussain}, \binits{S.}},
\bauthor{\bsnm{Rezazadegan}, \binits{D.}},
\bauthor{\bsnm{Liu}, \binits{L.}},
\bauthor{\bsnm{Ghavamzadeh}, \binits{M.}},
\bauthor{\bsnm{Fieguth}, \binits{P.}},
\bauthor{\bsnm{Cao}, \binits{X.}},
\bauthor{\bsnm{Khosravi}, \binits{A.}},
\bauthor{\bsnm{Acharya}, \binits{U.R.}}, \betal:
\batitle{A review of uncertainty quantification in deep learning: {T}echniques, applications and challenges}.
\bjtitle{Inf. Fusion}
\bvolume{76},
\bfpage{243}--\blpage{297}
(\byear{2021})
\end{barticle}
\endbibitem

\bibitem{wu2022integrated}
\begin{botherref}
\oauthor{\bsnm{Wu}, \binits{J.}},
\oauthor{\bsnm{Guo}, \binits{Y.}},
\oauthor{\bsnm{Deng}, \binits{C.}},
\oauthor{\bsnm{Zhang}, \binits{A.}},
\oauthor{\bsnm{Qiao}, \binits{H.}},
\oauthor{\bsnm{Lu}, \binits{Z.}},
\oauthor{\bsnm{Xie}, \binits{J.}},
\oauthor{\bsnm{Fang}, \binits{L.}},
\oauthor{\bsnm{Dai}, \binits{Q.}}:
An integrated imaging sensor for aberration-corrected 3{D} photography.
Nature,
62--71
(2022)
\end{botherref}
\endbibitem

\bibitem{gao2014single}
\begin{barticle}
\bauthor{\bsnm{Gao}, \binits{L.}},
\bauthor{\bsnm{Liang}, \binits{J.}},
\bauthor{\bsnm{Li}, \binits{C.}},
\bauthor{\bsnm{Wang}, \binits{L.V.}}:
\batitle{Single-shot compressed ultrafast photography at one hundred billion frames per second}.
\bjtitle{Nature}
\bvolume{516}(\bissue{7529}),
\bfpage{74}--\blpage{77}
(\byear{2014})
\end{barticle}
\endbibitem

\bibitem{altaqui2021mantis}
\begin{barticle}
\bauthor{\bsnm{Altaqui}, \binits{A.}},
\bauthor{\bsnm{Sen}, \binits{P.}},
\bauthor{\bsnm{Schrickx}, \binits{H.}},
\bauthor{\bsnm{Rech}, \binits{J.}},
\bauthor{\bsnm{Lee}, \binits{J.-W.}},
\bauthor{\bsnm{Escuti}, \binits{M.}},
\bauthor{\bsnm{You}, \binits{W.}},
\bauthor{\bsnm{Kim}, \binits{B.J.}},
\bauthor{\bsnm{Kolbas}, \binits{R.}},
\bauthor{\bsnm{O’Connor}, \binits{B.T.}}, \betal:
\batitle{Mantis shrimp--inspired organic photodetector for simultaneous hyperspectral and polarimetric imaging}.
\bjtitle{Sci. Adv.}
\bvolume{7}(\bissue{10}),
\bfpage{3196}
(\byear{2021})
\end{barticle}
\endbibitem

\bibitem{shi2020pre}
\begin{barticle}
\bauthor{\bsnm{Shi}, \binits{W.}},
\bauthor{\bsnm{Koo}, \binits{D.E.}},
\bauthor{\bsnm{Kitano}, \binits{M.}},
\bauthor{\bsnm{Chiang}, \binits{H.J.}},
\bauthor{\bsnm{Trinh}, \binits{L.A.}},
\bauthor{\bsnm{Turcatel}, \binits{G.}},
\bauthor{\bsnm{Steventon}, \binits{B.}},
\bauthor{\bsnm{Arnesano}, \binits{C.}},
\bauthor{\bsnm{Warburton}, \binits{D.}},
\bauthor{\bsnm{Fraser}, \binits{S.E.}}, \betal:
\batitle{Pre-processing visualization of hyperspectral fluorescent data with spectrally encoded enhanced representations}.
\bjtitle{Nat. Commun.}
\bvolume{11}(\bissue{1}),
\bfpage{726}
(\byear{2020})
\end{barticle}
\endbibitem

\bibitem{wu2021iterative}
\begin{barticle}
\bauthor{\bsnm{Wu}, \binits{J.}},
\bauthor{\bsnm{Lu}, \binits{Z.}},
\bauthor{\bsnm{Jiang}, \binits{D.}},
\bauthor{\bsnm{Guo}, \binits{Y.}},
\bauthor{\bsnm{Qiao}, \binits{H.}},
\bauthor{\bsnm{Zhang}, \binits{Y.}},
\bauthor{\bsnm{Zhu}, \binits{T.}},
\bauthor{\bsnm{Cai}, \binits{Y.}},
\bauthor{\bsnm{Zhang}, \binits{X.}},
\bauthor{\bsnm{Zhanghao}, \binits{K.}}, \betal:
\batitle{Iterative tomography with digital adaptive optics permits hour-long intravital observation of {3D} subcellular dynamics at millisecond scale}.
\bjtitle{Cell}
\bvolume{184}(\bissue{12}),
\bfpage{3318}--\blpage{3332}
(\byear{2021})
\end{barticle}
\endbibitem

\bibitem{gehm2007single}
\begin{barticle}
\bauthor{\bsnm{Gehm}, \binits{M.E.}},
\bauthor{\bsnm{John}, \binits{R.}},
\bauthor{\bsnm{Brady}, \binits{D.J.}},
\bauthor{\bsnm{Willett}, \binits{R.M.}},
\bauthor{\bsnm{Schulz}, \binits{T.J.}}:
\batitle{Single-shot compressive spectral imaging with a dual-disperser architecture}.
\bjtitle{Opt. Express}
\bvolume{15}(\bissue{21}),
\bfpage{14013}--\blpage{14027}
(\byear{2007})
\end{barticle}
\endbibitem

\bibitem{cao2011prism}
\begin{barticle}
\bauthor{\bsnm{Cao}, \binits{X.}},
\bauthor{\bsnm{Du}, \binits{H.}},
\bauthor{\bsnm{Tong}, \binits{X.}},
\bauthor{\bsnm{Dai}, \binits{Q.}},
\bauthor{\bsnm{Lin}, \binits{S.}}:
\batitle{A prism-mask system for multispectral video acquisition}.
\bjtitle{IEEE T. Pattern Anal.}
\bvolume{33}(\bissue{12}),
\bfpage{2423}--\blpage{2435}
(\byear{2011})
\end{barticle}
\endbibitem

\bibitem{kim20123d}
\begin{barticle}
\bauthor{\bsnm{Kim}, \binits{M.H.}},
\bauthor{\bsnm{Harvey}, \binits{T.A.}},
\bauthor{\bsnm{Kittle}, \binits{D.S.}},
\bauthor{\bsnm{Rushmeier}, \binits{H.}},
\bauthor{\bsnm{Dorsey}, \binits{J.}},
\bauthor{\bsnm{Prum}, \binits{R.O.}},
\bauthor{\bsnm{Brady}, \binits{D.J.}}:
\batitle{3{D} imaging spectroscopy for measuring hyperspectral patterns on solid objects}.
\bjtitle{ACM Trans. Graph.}
\bvolume{31}(\bissue{4}),
\bfpage{1}--\blpage{11}
(\byear{2012})
\end{barticle}
\endbibitem

\bibitem{lin2014spatial}
\begin{barticle}
\bauthor{\bsnm{Lin}, \binits{X.}},
\bauthor{\bsnm{Liu}, \binits{Y.}},
\bauthor{\bsnm{Wu}, \binits{J.}},
\bauthor{\bsnm{Dai}, \binits{Q.}}:
\batitle{Spatial-spectral encoded compressive hyperspectral imaging}.
\bjtitle{ACM Trans. Graph.}
\bvolume{33}(\bissue{6}),
\bfpage{1}--\blpage{11}
(\byear{2014})
\end{barticle}
\endbibitem

\bibitem{ma2014acquisition}
\begin{barticle}
\bauthor{\bsnm{Ma}, \binits{C.}},
\bauthor{\bsnm{Cao}, \binits{X.}},
\bauthor{\bsnm{Tong}, \binits{X.}},
\bauthor{\bsnm{Dai}, \binits{Q.}},
\bauthor{\bsnm{Lin}, \binits{S.}}:
\batitle{Acquisition of high spatial and spectral resolution video with a hybrid camera system}.
\bjtitle{Int. J Comput. Vision}
\bvolume{110},
\bfpage{141}--\blpage{155}
(\byear{2014})
\end{barticle}
\endbibitem

\bibitem{lin2014dual}
\begin{barticle}
\bauthor{\bsnm{Lin}, \binits{X.}},
\bauthor{\bsnm{Wetzstein}, \binits{G.}},
\bauthor{\bsnm{Liu}, \binits{Y.}},
\bauthor{\bsnm{Dai}, \binits{Q.}}:
\batitle{Dual-coded compressive hyperspectral imaging}.
\bjtitle{Opt. Lett.}
\bvolume{39}(\bissue{7}),
\bfpage{2044}--\blpage{2047}
(\byear{2014})
\end{barticle}
\endbibitem

\bibitem{golub2016compressed}
\begin{barticle}
\bauthor{\bsnm{Golub}, \binits{M.A.}},
\bauthor{\bsnm{Averbuch}, \binits{A.}},
\bauthor{\bsnm{Nathan}, \binits{M.}},
\bauthor{\bsnm{Zheludev}, \binits{V.A.}},
\bauthor{\bsnm{Hauser}, \binits{J.}},
\bauthor{\bsnm{Gurevitch}, \binits{S.}},
\bauthor{\bsnm{Malinsky}, \binits{R.}},
\bauthor{\bsnm{Kagan}, \binits{A.}}:
\batitle{Compressed sensing snapshot spectral imaging by a regular digital camera with an added optical diffuser}.
\bjtitle{Appl. Opt.}
\bvolume{55}(\bissue{3}),
\bfpage{432}--\blpage{443}
(\byear{2016})
\end{barticle}
\endbibitem

\bibitem{wang2018computational}
\begin{barticle}
\bauthor{\bsnm{Wang}, \binits{P.}},
\bauthor{\bsnm{Menon}, \binits{R.}}:
\batitle{Computational multispectral video imaging}.
\bjtitle{J. Opt. Soc. Am.}
\bvolume{35}(\bissue{1}),
\bfpage{189}--\blpage{199}
(\byear{2018})
\end{barticle}
\endbibitem

\bibitem{mu2019compact}
\begin{barticle}
\bauthor{\bsnm{Mu}, \binits{T.}},
\bauthor{\bsnm{Han}, \binits{F.}},
\bauthor{\bsnm{Bao}, \binits{D.}},
\bauthor{\bsnm{Zhang}, \binits{C.}},
\bauthor{\bsnm{Liang}, \binits{R.}}:
\batitle{Compact snapshot optically replicating and remapping imaging spectrometer {(ORRIS)} using a focal plane continuous variable filter}.
\bjtitle{Opt. Lett.}
\bvolume{44}(\bissue{5}),
\bfpage{1281}--\blpage{1284}
(\byear{2019})
\end{barticle}
\endbibitem

\bibitem{mcclung2020snapshot}
\begin{barticle}
\bauthor{\bsnm{McClung}, \binits{A.}},
\bauthor{\bsnm{Samudrala}, \binits{S.}},
\bauthor{\bsnm{Torfeh}, \binits{M.}},
\bauthor{\bsnm{Mansouree}, \binits{M.}},
\bauthor{\bsnm{Arbabi}, \binits{A.}}:
\batitle{Snapshot spectral imaging with parallel metasystems}.
\bjtitle{Sci. Adv.}
\bvolume{6}(\bissue{38}),
\bfpage{7646}
(\byear{2020})
\end{barticle}
\endbibitem

\bibitem{williams2019grayscale}
\begin{barticle}
\bauthor{\bsnm{Williams}, \binits{C.}},
\bauthor{\bsnm{Gordon}, \binits{G.S.}},
\bauthor{\bsnm{Wilkinson}, \binits{T.D.}},
\bauthor{\bsnm{Bohndiek}, \binits{S.E.}}:
\batitle{Grayscale-to-color: scalable fabrication of custom multispectral filter arrays}.
\bjtitle{ACS Photonics}
\bvolume{6}(\bissue{12}),
\bfpage{3132}--\blpage{3141}
(\byear{2019})
\end{barticle}
\endbibitem

\bibitem{zhang2023handheld}
\begin{barticle}
\bauthor{\bsnm{Zhang}, \binits{W.}},
\bauthor{\bsnm{Suo}, \binits{J.}},
\bauthor{\bsnm{Dong}, \binits{K.}},
\bauthor{\bsnm{Li}, \binits{L.}},
\bauthor{\bsnm{Yuan}, \binits{X.}},
\bauthor{\bsnm{Pei}, \binits{C.}},
\bauthor{\bsnm{Dai}, \binits{Q.}}:
\batitle{Handheld snapshot multi-spectral camera at tens-of-megapixel resolution}.
\bjtitle{Nat. Commun.}
\bvolume{14}(\bissue{1}),
\bfpage{5043}
(\byear{2023})
\end{barticle}
\endbibitem

\bibitem{yuan2023super}
\begin{barticle}
\bauthor{\bsnm{Yuan}, \binits{L.}},
\bauthor{\bsnm{Song}, \binits{Q.}},
\bauthor{\bsnm{Liu}, \binits{H.}},
\bauthor{\bsnm{Heggarty}, \binits{K.}},
\bauthor{\bsnm{Cai}, \binits{W.}}:
\batitle{Super-resolution computed tomography imaging spectrometry}.
\bjtitle{Photonics Res.}
\bvolume{11}(\bissue{2}),
\bfpage{212}--\blpage{224}
(\byear{2023})
\end{barticle}
\endbibitem

\end{thebibliography}
\end{document}